\documentclass[twocolumn,preprintnumbers,amsmath,amssymb]{revtex4}
\usepackage{amssymb}
\usepackage{latexsym}
\usepackage{epsfig}
\usepackage{color}

\usepackage{color}

\begin{document}

\title{Energy definition and dark energy: a thermodynamic analysis}

\author{H. Moradpour$^1$\footnote{hn.moradpour@gmail.com}, J. P.
Morais Gra\c ca$^{2}$\footnote{jpmorais@gmail.com}, I. P.
Lobo$^{2}$\footnote{iarley\_lobo@fisica.ufpb.br}, I. G.
Salako$^{3}$\footnote{inessalako@gmail.com}}
\address{$^1$ Research Institute for Astronomy and Astrophysics of Maragha
(RIAAM), P.O. Box 55134-441, Maragha, Iran\\
$^{2}$ Departamento de F\'{i}sica, Universidade Federal da
Para\'{i}ba, Caixa Postal 5008, CEP 58051-970, Jo\~{a}o Pessoa,
PB, Brazil\\
$^3$ Institut de Math\'ematiques et de Sciences Physiques (IMSP)
01 BP 613 Porto-Novo, B\'enin}

\pacs{ }

\begin{abstract}
\noindent Accepting the Komar mass definition of a source with
energy-momentum tensor $T_{\mu\nu}$, and using the thermodynamic
pressure definition, we find a relaxed energy-momentum
conservation law. Thereinafter, we study some cosmological
consequences of the obtained energy-momentum conservation law. It
has been found out that the dark sectors of cosmos are unifiable
into one cosmic fluid in our setup. While this cosmic fluid impels
the universe to enter an accelerated expansion phase, it may
even show a baryonic behavior by itself during the cosmos
evolution. Indeed, in this manner, while $T_{\mu\nu}$ behaves
baryonically, some parts of it, namely $T_{\mu\nu}(e)$ which is
satisfying the ordinary energy-momentum conservation law, are
responsible for the current accelerated expansion.
\end{abstract}

\maketitle
\section{Introduction}

Friedmann equations, the ordinary energy-momentum conservation
law (OCL) (or the continuity equation) and its compatibility with the Bianchi Identity (BI) are the backbone of the
standard cosmology which forms the foundation of our understanding
of cosmos \cite{roos}. Since on scales larger than about $100$
Megaparsecs, cosmos is homogeneous and isotropic \cite{roos}, the FRW
metric is a suitable metric to study the cosmic evolution
\cite{roos}. Relations between thermodynamics and Friedmann
equations have  been studied in various setups which help us in
getting more close to the thermodynamic origin of spacetime,
gravity and related topics
\cite{Hay22,Hay2,Bak,j20,j200,j2,caiwork,j3,j4,sheyw1,sheyw2,ufl,em,md,non13,non20,plb,mr,mpla}.

A thermodynamic analysis can also lead to a better understanding
of the origin of dark energy, responsible for the current
accelerated universe
\cite{em,md,non13,non20,mpla,mr,msrw,pavr,pavr1,pavr11,pavr12,pavonf0,pavonf,prdn,S1,S2,S3,rafael,mae}.
In fact, there are thermodynamic and holographic approaches
claiming that the cosmos expansion can be explained as an emergent
phenomenon
\cite{non20,km1,km2,km3,km4,km5,km6,km7,km8,km9,km10,km11,km12,km13}.
Two key points in these approaches are the definition of energy
and the form of the energy-momentum conservation law
\cite{plb,mpla,km1,km2,km3,km4,km7}, and indeed, their results are
so sensitive to the energy definitions that have been employed
\cite{mpla}.

In order to make the discussion clearer, consider a flat FRW
universe with scale factor $a(t)$ \cite{roos}

\begin{eqnarray}\label{frw}
ds^{2}=-dt^{2}+a^{2}\left( t\right) \left[
dr^{2}+r^{2}d\Omega^{2}\right],
\end{eqnarray}

\noindent filled by a prefect fluid source with energy density $\rho$ and
pressure $p$. Einstein field equations and energy-momentum
conservation law (or equally the Bianchi identity) lead to

\begin{eqnarray}\label{fried1}
&&H^2=\frac{8\pi}{3}\rho,\\
&&3H^2+2\dot{H}=-8\pi p,\nonumber\\
&&\dot{\rho}+3H(\rho+p)=0,\nonumber
\end{eqnarray}

\noindent forming the cornerstone of standard cosmology. Although,
the third equation (OCL) is in full agreement with our
observations on the cosmic fluid in matter and radiation dominated eras
\cite{roos,wein}, its validity for the current accelerated cosmos
is questionable \cite{prl} which may encourage us to consider the
relaxed types of OCL to describe the current
universe \cite{genr,non20}. Combining the Friedmann equations with
each other, we get

\begin{eqnarray}\label{ray}
\dot{H}=-4\pi(\rho+p),
\end{eqnarray}

\noindent which can finally be added to the time derivative of the
first Friedmann equation to reach at OCL. Therefore, although the
third equation of~(\ref{fried1}) is generally obtained from the
conservation law ($T_{\mu\nu}^{\ \ ;\nu}=0$), one can easily find
it by only using the Friedmann equations, a result which is the
direct consequence of the Einstein field equations satisfying the
Bianchi identity. It means that two equations of the above set of
equations~(\ref{fried1}) are indeed enough to study the cosmos,
and the third equation will be automatically valid and inevitable
in this situation.

From the viewpoint of thermodynamics, if OCL is valid, then by
applying the thermodynamics laws to the apparent horizon, as the
proper causal boundary located at \cite{Hay22,Hay2,Bak}

\begin{eqnarray}\label{ah}
\tilde{r}_A=a(t)r_A=\frac{1}{H},
\end{eqnarray}

\noindent where $r_A$ is the co-moving radius of apparent horizon,
one can reach the Friedmann equations~(\ref{fried1})
\cite{caiwork,sheyw1,sheyw2,ufl,em,md,non13,non20,j20,j200,j2,j3,j4,plb}.
Indeed, the validity of OCL is crucial, and its relaxation leads
to modification of the Friedmann equations \cite{non20,em,md,plb}.
These results propose that the cosmos is so close to its
thermodynamic equilibrium state
\cite{mr,msrw,pavr,pavr1,pavr11,pavr12,pavonf0,pavonf,prdn,mpla},
motivating us to be more confident on thermodynamical quantities,
such as pressure, to study the cosmos \cite{mpla}.

For a FRW background filled by a fluid whose energy density $\rho$
is at most a function of time, we may reach

\begin{eqnarray}\label{Ene1}
E=\int T_{\alpha\beta}u^\alpha u^\beta dV=\int \rho dV=\rho V,
\end{eqnarray}

\noindent as the total energy confined to the volume $V$ and felt
by a co-moving observer with four velocity
$u^\alpha=\delta^\alpha_t$. Eq.~(\ref{Ene1}) is fully compatible
with the combination of the Misner-Sharp mass, thermodynamics laws
and standard cosmology~(\ref{fried1})
\cite{caiwork,sheyw1,sheyw2,ufl,em,md,non13,non20,j20,j200,j2,j3,j4,plb}.

Now, if we write $V=V_0 a^3$, where $V_0$ is the co-moving volume
\cite{roos}, use Eq.~(\ref{Ene1}), and the thermodynamic pressure definition

\begin{eqnarray}\label{pres1}
p=-\Big(\frac{\partial E}{\partial V}\Big),
\end{eqnarray}

\noindent then we easily get OCL. In fact, it shows that the form of the conservation law depends on the energy definition. The above arguments also
motivated the authors of Ref.~\cite{mpla} to use various energy
definitions such as the Komar mass \cite{komar0,komar1} and

\begin{eqnarray}\label{fried2}
&&H^2=\frac{8\pi}{3}\rho,\\
&&3H^2+2\dot{H}=8\pi \Big(\frac{\partial E}{\partial
V}\Big),\nonumber\\
&&\dot{\rho}+3H(\rho+p)=0,\nonumber
\end{eqnarray}

\noindent instead of Eqs.~(\ref{fried1}), in order to study the
relations between energy definitions, thermodynamics laws and the
cosmic evolution, which led to a model for the cosmic fluid
consistent with observations \cite{mpla}.

The Einstein field equations also admit another mass definition,
namely the Komar mass \cite{komar0,komar1,wein}

\begin{eqnarray}\label{tkm}
\mathcal{M}=2\int (T_{\mu\nu}-\frac{1}{2}Tg_{\mu\nu}) u^\mu u^\nu
dV=(\rho+3p)V,
\end{eqnarray}

\noindent which clearly differs from~(\ref{Ene1}) unless we have a
pressureless fluid \cite{km8,wein}. Although this energy
definition has originally been obtained by using the Einstein
field equations \cite{wein}, it is in accordance with various
thermodynamic and holographic approaches
\cite{km1,km2,km3,km4,km5,km6,km7,km8,km9,km10,km11,km12,km13}. In
fact, if we accept this energy definition as a basic equation and
not a secondary equation \cite{km1,km2,km3,km4}, then there are
various thermodynamic and holographic approaches employed to get
the Friedmann and gravitational field equations in various
theories by using the Komar mass definition
\cite{km1,km2,km3,km4,km5,km6,km7,km8,km9,km10,km11,km12,km13}.

Now, inserting the Komar energy definition into Eq.~(\ref{pres1})
and using $V=V_0 a^3$, one can easily reach a new conservation law
as

\begin{eqnarray}\label{memc}
\dot{\rho}+3(\dot{p}+3Hp)+3H(\rho+p)=0,
\end{eqnarray}

\noindent met by Komar mass, and we call it the Komar conservation
law (KCL). For the pressureless systems, everything is consistent
due to the fact that the above addressed energy definitions are
equal. Similar types of this energy-momentum conservation law have
also been obtained in theories which include a non-minimal
coupling between the geometry and matter fields
\cite{rastall,genr,plb}.

One basic consequence of the Komar mass is that not only $\rho$, but
both of $\rho$ and $p$ participate in building the system energy
which leads to appropriate models for dark sectors of cosmos, if
one uses Eq.~(\ref{fried2}) to model the cosmos \cite{mpla}. On
the other hand, Eq.~(\ref{memc}) claims that if we use the Komar
mass, then OCL and thus Eq.~(\ref{fried2}) are not valid and
another modified Friedmann equations should be employed to
describe the cosmos.

In the general relativity (GR) framework, where $G_{\mu\nu}=8\pi T_{\mu\nu}$, the
satisfaction of BI by the Einstein tensor ($G_\mu^\nu$) is
equivalent to the satisfaction of OCL by $T_\mu^\nu$. We saw that
the form of conservation law depends on the energy
definition~(\ref{pres1}), and indeed, if the Komar mass notion is
used, then one can easily find that $T_\mu^\nu$ should satisfy KCL
instead of OCL. Thus, in the GR framework, there is an inconsistency
between the notions of BI (or equally OCL) and the Komar mass. This inconsistency together with the results of recent observations \cite{prl}, admitting the break-down of OCL in the current accelerated era, motivate us to modify GR, and thus, the Friedmann equations by
modifying the matter side of the field equations in the manner in
which the modified matter part meets OCL, in agreement with the
satisfaction of BI by the geometrical part. In summary, we think
that OCL is very restrictive constraint on $T_\mu^\nu$. Probably, OCL should not be applied to
whole of $T_\mu^\nu$, and it should only be applied to some parts of $T_\mu^\nu$, the
parts which may modify GR in a compatible way with the current accelerated cosmos \cite{prl,genr,non20}.

There is also an elegant consistency between Eqs.~(\ref{Ene1})
and~(\ref{fried1}), i.e. $\rho$ is the energy density in
Eq.~(\ref{Ene1}), and it also appears in the first Friedmann
equation. On the other hand, Eq.~(\ref{tkm}) implies that the
quantity ($\rho+3p$) plays the role of energy density, while it is
not present in the first Friedmann equation. In fact, if we define
$\rho_e\equiv\rho+3p$, then Eq.~(\ref{memc}) can be rewritten as

\begin{eqnarray}\label{memc1}
\dot{\rho}_e+3H(\rho_e+p)=0,
\end{eqnarray}

\noindent equivalent to a hypothetical fluid with an effective
energy-momentum tensor of $T_\mu^\nu(e)=diag(-\rho_e,p,p,p)$,
meeting OCL (see~\cite{dmlhc} for a debate on effective
energy-momentum tensor). In this manner, Eq.~(\ref{Ene1}) for
$T_\mu^\nu(e)$ will be equal to Eq.~(\ref{tkm}) for $T_\mu^\nu$.
Indeed, the above arguments (specially Eq.~(\ref{memc})) claim
that, even in the Einstein framework, OCL~(\ref{fried1}) may not
be valid, if Komar mass is employed. In this situation, because
OCL is the backbone of the Friedmann equations, we may conclude
that the Friedmann equations~(\ref{fried1}) should be modified.

In summary, \textit{$i$) The energy definition and
the form of energy-momentum conservation law play crucial roles in
getting the Friedmann equations in various theories
\cite{caiwork,sheyw1,sheyw2,ufl,em,md,non13,non20,j20,j200,j2,j3,j4,mpla,plb}.
$ii$) We also showed the form of energy-momentum conservation law
depends on the energy definition, and if one uses the Komar mass,
then KCL is obtained instead of OCL. $iii$) The Komar mass is the
backbone of some holographic and thermodynamic attempts to obtain
field equations of various gravitational theory
\cite{km1,km2,km3,km4,km5,km6,km7,km8,km9,km10,km11,km12,km13}
motivating us to use KCL instead of OCL. $iv$) Observational data
admit the break-down of OCL in the current accelerated era
\cite{prl}. It also motivates us to use the relaxed forms of OCL in
order to describe the current universe.} On the other
hand, there are also various models for the current accelerated
universe satisfying OCL \cite{roos}. Therefore, it seems that some
inconsistency exists between the various definitions of
energy, continuity and Friedmann equations. One may conclude that
the fluid obtained from the observation may not be the real
cosmic fluid, and it indeed represents some less well-known or even
unknown parts of the energy source which become
important, effective and tangible in the current cosmos. We want to
say that the real cosmic fluid may satisfy conservation laws such as
KCL instead of OCL, while some of its parts, responsible for the
current accelerated expansion, may satisfy OCL.

As we mentioned, based on Eq.~(\ref{tkm}), one may conclude that
pressure has some contribution to the energy density of system.
The question arisen here is what if this contribution can describe
the current accelerated universe? In other words, are there
combinations of $\rho$ and $p$ of a baryonic source which can play
the role of dark energy? Here, we are going to find out the
answers of the mentioned questions by studying some consequences
of accepting Eq.~(\ref{pres1}) and the Komar mass definition, and
thus Eq.~(\ref{memc}), in cosmological setups. In the next
section, using thermodynamics laws, Bekenstein entropy and
Eq.~(\ref{memc}), we address some models for dark energy. The
third section is devoted to a summary and concluding remarks.
Throughout this work, the unit of $c=\hbar=G=k_B=1$, where $k_B$
denotes the Boltzmann constant, has been employed.

\section{Komar definition of energy and the Friedmann equations}

For an energy-momentum source with $T_a^b=diag(-\rho,p,p,p)$, the
amount of energy crossing the apparent horizon is evaluated as
\cite{caiwork}

\begin{eqnarray}\label{uf1}
\delta Q^m=A(T^b_a\partial_b \tilde{r}_A + W\partial_a
\tilde{r}_A)dx^a,
\end{eqnarray}

\noindent where $A=4\pi\tilde{r}_A^2=A_0 a^2$ and
$W=\frac{\rho-p}{2}$ denote the horizon area and the work density,
respectively. $A_0=4\pi r_A^2$ is also the co-moving area. After
some calculations, one obtains \cite{caiwork,non13,em,md,plb}

\begin{eqnarray}\label{ufl2}
\delta Q^m=-3VH(\rho+p)dt,
\end{eqnarray}

\noindent combined with the Clausius relation ($TdS_A=-\delta
Q^m$), to get \cite{caiwork,em}

\begin{eqnarray}\label{ufl3}
3VH(\rho+p)dt=\frac{H}{2\pi}dS_A,
\end{eqnarray}

\noindent where the additional minus sign in the Clausius relation
is due to the direction of energy flux \cite{caiwork,em}. Here,
$S_A$ is the horizon entropy and the Cai-Kim temperature
($T=\frac{H}{2\pi}$) is used to reach this equation
\cite{caiwork,em}. The role of energy-momentum conservation law in
deriving the Friedmann equations is now completely clear. The
effects of considering various approximations and temperatures
have been studied in Ref.~\cite{em}.

\subsection*{Modified Friedmann equations}

Here, using some simple examples, we are going to
show the effects of considering KCL in modifying the standard
cosmology.

Case $i$) If OCL is valid, then Eq.~(\ref{ufl3})
is reduced to

\begin{eqnarray}\label{ufl4}
d\rho=-\frac{3H^4}{8\pi^2}dS_A,
\end{eqnarray}

\noindent covering the first Friedmann equation whenever
$S=\frac{A}{4}$. In this situation, Eq.~(\ref{memc}) implies that

\begin{eqnarray}\label{ufl5}
\dot{p}+3Hp=0\rightarrow p(a)=p_0a^{-3},
\end{eqnarray}

\noindent where $p_0$ is the integration constant and we
considered $a_0=1$ ($a_0$ is the current value of the scale
factor). In this manner, for $S_A=\frac{A}{4}$, we reach

\begin{eqnarray}\label{fried3}
&&H^2=\frac{8\pi}{3}\rho,\\
&&3H^2+2\dot{H}=-8\pi p_0a^{-3},\nonumber\\
&&\dot{\rho}+3H(\rho+p_0a^{-3})=0,\nonumber
\end{eqnarray}

\noindent instead of Eqs.~(\ref{fried1}). It is easy to check that
for power law regimes in which $\rho\propto a^\beta$, conservation
law leads to $p_0=0$ and $\beta=-3$, which is nothing but the dust
source, in full agreement with our previous results indicating that
for a pressureless source everything is consistent.
The geometrical parts of the two first equations of~(\ref{fried3})
address us to the Einstein tensor. It is in fact the direct result
of attributing the Bekenstein entropy to the horizon \cite{caiwork}.
Therefore, mathematically, since the Einstein tensor obeys BI, the
appeared $\rho$ and $p$ should also satisfy OCL, an expectation in
full agreement with Eq.~(\ref{ufl5}) and thus the third line of Eqs.~(\ref{fried3}).

\begin{figure}[ht]
\centering
\includegraphics[width=2in, height=2in]{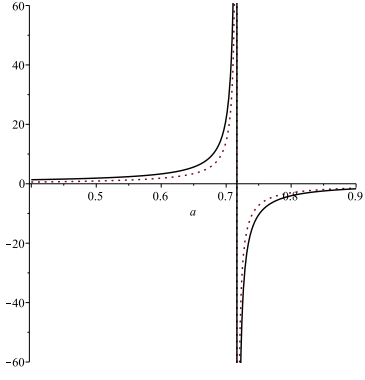}
\caption{\label{fig1} Here, $w_0=-1$, dot and solid lines
represent $w$ and $q$, respectively.}
\end{figure}

However, the general solution of the obtained conservation
law~(\ref{fried3}) is

\begin{eqnarray}\label{cons1}
\rho(a)=a^{-3}[c-3p_0\ln(a)],
\end{eqnarray}

\noindent where $c$ is the integration constant. It clearly covers
the ordinary matter era whenever $p_0=0$. From Eqs.~(\ref{ufl5})
and~(\ref{cons1}), we reach

\begin{eqnarray}\label{cons2}
p=p_0,\ \ \rho=c,
\end{eqnarray}

\noindent for the current era. Thus, if we have $p_0=w_0 c$ and
$c=\Lambda>0$, where $\Lambda$ and $w_0$ denote the current values
of the energy density and the equation of state (EoS) of dark
energy, respectively, then the obtained fluid may theoretically
generate the current accelerating universe whenever
$w_0\leq-\frac{2}{3}$. Now, we can write

\begin{eqnarray}\label{cons2}
p(a)=w_0\Lambda a^{-3},\ \ \rho(a)=\Lambda a^{-3}[1-3w_0\ln(a)].
\end{eqnarray}

\noindent In this manner, EoS ($w$) and the deceleration parameter
($q$) are evaluated as

\begin{eqnarray}\label{cons2}
&&w\equiv\frac{p}{\rho}=\frac{w_0}{1-3w_0\ln(a)},\\
&&q\equiv-1-\frac{\dot{H}}{H^2}=\frac{1}{2}(1+\frac{3w_0}{1-3w_0\ln(a)}).\nonumber
\end{eqnarray}

\noindent At the $a\rightarrow0$ limit, independent of the value
of $w_0$, we have $w\rightarrow0$ and $q\rightarrow\frac{1}{2}$.
Moreover, one can easily see that, at the $a\rightarrow1$ limit,
the consistent values with observations for $w$ and $q$ are also
obtainable, depending on the value of $w_0$. Despite this
compatibility, there is a singularity at the behavior of $q$ and
$w$ for $w_0<0$, located at $a=e^{\frac{1}{3w_0}}$, meaning that
this solution may be at most useable to study the
$a>e^{\frac{1}{3w_0}}$ era. But, as it is obvious from
Fig.~(\ref{fig1}), it does not lead to suitable $q$ and $w$ even
for $a>e^{\frac{1}{3w_0}}$. Therefore, as we previously saw, only
the $w_0=0$ case can meet all of the desired expectations, i.e. it
is in agreement with the Friedmann equations, the both addressed
energy definitions, and thus the matter dominated era.

Case $ii$) If $T_\mu^\nu(e)$, satisfying
Eq.~(\ref{memc1}), is used instead of $T_\mu^\nu$ in order to
evaluate $\delta Q^m$~(\ref{uf1}), and we follow the Cai-Kim
approach \cite{caiwork,em,plb,md,non13,non20}, then some
calculations lead to

\begin{eqnarray}\label{fried4}
&&H^2=\frac{8\pi}{3}\rho_e,\\
&&3H^2+2\dot{H}=-8\pi p,\nonumber\\
&&\dot{\rho}_e+3H(\rho_e+p)=0,\nonumber
\end{eqnarray}

\noindent whenever $S_A=\frac{A}{4}$. These equations would also be
obtainable by modifying the Einstein field equations as
$G_{\mu\nu}=8\pi T_{\mu\nu}(e)$, in agreement with
attributing the Bekenstein entropy to the apparent horizon (see the
paragraph after Eqs.~(\ref{fried3})). Now, considering the simple
case $w_e\equiv\frac{p}{\rho_e}$ and using the above equations, we
can easily find $w_e=\frac{w}{1+3w}$. This result indicates that a
baryonic source, with $w>0$, cannot lead to a hypothetical fluid
with $w_e<0$ in cosmos, meaning that the obtained modified Friedmann
equations cannot describe the current accelerated universe.


Case $iii$) Now, bearing Eq.~(\ref{memc}) in mind,
and defining an effective pressure as $p_e=\beta p$, one reaches

\begin{eqnarray}\label{memcf1}
\dot{\rho}+\frac{3}{\beta}[\dot{p}_e+(4-\beta)Hp_e]+3H(\rho+p_e)=0,
\end{eqnarray}

\noindent reduced to

\begin{eqnarray}\label{memcf2}
\dot{\rho}+3H(\rho+p_e)=0,
\end{eqnarray}

\noindent whenever we have

\begin{eqnarray}\label{memcf3}
\dot{p}_e+(4-\beta)Hp_e=0\Rightarrow p_e=p_e^0a^{\beta-4},
\end{eqnarray}

\noindent where $p_e^0$ is the integration constant. Inserting
this result into Eq.~(\ref{memcf2}), it is easy to obtain

\begin{eqnarray}\label{memcf4}
\rho=\frac{c}{a^3}-3\frac{p_e^0}{\beta-1}a^{\beta-4}.
\end{eqnarray}

\noindent Here, $c$ is an integration constant, and it is worthwhile
to mention that the ordinary matter and radiation dominated eras are
recovered at the limits of $p_e^0=0$ and $c,\beta=0$, respectively,
meaning that the obtained fluid can unify these eras into one model.
In this manner, we can argue that although $T_{\mu\nu}$ is the true
energy-momentum tensor, an effective tensor of
$T_\mu^\nu(e)=diag(-\rho,p_e,p_e,p_e)$ should be considered, which
may address us to a modified gravitational equation of the form
$G_{\mu\nu}=8\pi T_{\mu\nu}(e)$. Now, combining Eqs.~(\ref{memcf3})
and~(\ref{memcf4}) with each other, one finds

\begin{eqnarray}\label{memcf5}
w_e\equiv\frac{p_e}{\rho}=\beta
w=\frac{(\beta-1)p_e^0a^3}{(\beta-1)ca^{4-\beta}-3p_e^0a^3}.
\end{eqnarray}

\noindent For $\beta>1$ and $\beta<1$, we have $w_e\rightarrow0$
and $w_e\rightarrow(1-\beta)/3$, respectively, at the
$a\rightarrow0$ limit. Moreover, independent of the value of
$\beta$, one can see that
$w_e\rightarrow\frac{(\beta-1)p_e^0}{(\beta-1)c-3p_e^0}\equiv
w_0$, leading to $c=\frac{p_e^0(\beta-1+3w_0)}{(\beta-1)w_0}$,
whenever $a\rightarrow1$. Using the latest result and the
$1+z=a^{-1}$ relation, where $z$ is the redshift, we can finally
rewrite Eq.~(\ref{memcf5}) as

\begin{eqnarray}\label{memcf6}
w_e=\beta
w=\frac{w_0(\beta-1)}{(\beta-1+3w_0)(1+z)^{\beta-1}-3w_0}.
\end{eqnarray}

\begin{figure}[ht]
\centering
\includegraphics[width=2in, height=2in]{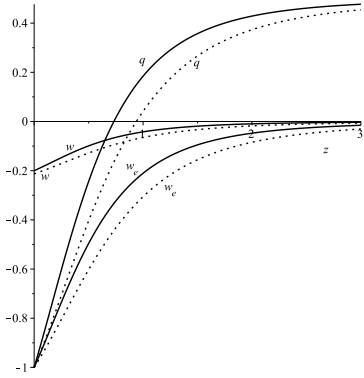}
\caption{\label{fig2} $w_e$, $w$ and $q$ versus $z$ when $w_0=-1$. Here, solid lines: $\beta=5$ and dot lines: $\beta=4.7$.}
\end{figure}

If $w_0<0$ and $c>0$, then for $\beta>1-3w_0$ and $p_e^0<0$ (or
equally $p_e<0$) we have $\rho>0$, $p<0$, and
$w_e(z\gg1)\rightarrow0$ meaning that this fluid may describe the
universe history from the matter dominated era to the current
accelerating phase. In this manner, if we either modify Einstein
equations as $G_{\mu\nu}=8\pi T_{\mu\nu}(e)$ or follow the Cai-Kim
recipe by using $T_{\mu\nu}(e)=diag(-\rho,p_e,p_e,p_e)$, then we
obtain

\begin{eqnarray}\label{fried5}
&&H^2=\frac{8\pi}{3}\rho,\\
&&3H^2+2\dot{H}=-8\pi p_e,\nonumber\\
&&\dot{\rho}+3H(\rho+p_e)=0.\nonumber
\end{eqnarray}

\noindent As before, two first equations imply that
$T_{\mu\nu}(e)$ should satisfy OCL, in full agreement with
Eq.~(\ref{memcf2}), and thus the third line of~(\ref{fried5}).
Calculations for the deceleration parameter lead to

\begin{eqnarray}\label{dec1}
q\equiv-1-\frac{\dot{H}}{H^2}=\frac{1}{2}(1+3w_e).
\end{eqnarray}

\noindent Suitable values of $\beta$ should be found by considering
the transition point at which $q_t=0$ and $w_e=-\frac{1}{3}\equiv
w_e^t$. In addition, simple calculations for the transition redshift
lead to

\begin{eqnarray}\label{trans1}
z_t=-1+(\frac{\beta-1+3w_0}{3w_0(2-\beta)})^{\frac{1}{1-\beta}},
\end{eqnarray}

\noindent indicating that $\beta$ should also meet the $\beta>2$
condition to reach positive meaningful solutions for $z_t$.
Because $w_0<w_e^t$, we have the relation $1-3w_0>2$ meaning that
the $\beta>2$ condition is in agreement with the previously
obtained condition ($\beta>1-3w_0$). Besides, although we can
obtain suitable values of $z_t$ for $1<\beta<1-3w_0$, in this
manner, negative amounts will be attained by $\rho$ which finally
reject this case. For other values of $\beta$, energy density will
grow much faster than the matter density, proportional to
$(1+z)^3$, by increasing $z$. Thus, only the $\beta>1-3w_0$ case
can be meaningful. In Fig.~(\ref{fig2}), deceleration parameter,
$w_e$ and $w$ have been plotted versus $z$ for some values of
$\beta$.

Hence, if we accept both the Komar mass definition and the
thermodynamic pressure notion, we can modify the standard
cosmology as~(\ref{fried5}) in which the solutions of the field
equations can theoretically unify both the matter dominated era
and the current accelerated universe. But, in this manner, since
$\beta>0$, the original source has a negative pressure
($p=\frac{p_e}{\beta}<0$). Therefore, it is useful to emphasize
that, despite the proper obtained results, since $p<0$, this
approach does not lead to a baryonic model for the energy source
during the cosmos evolution, i.e. a model with $0\leq p\leq
\frac{1}{3}\rho$.

\begin{figure}[ht]
\centering
\includegraphics[width=2in, height=2in]{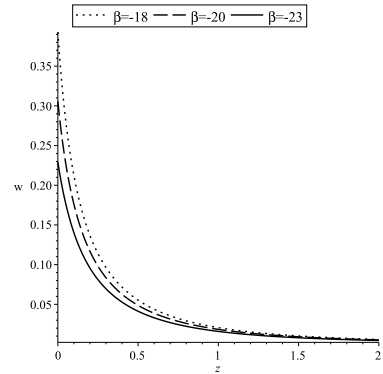}
\caption{\label{fig4} $w$ has been depicted versus $z$ for some values of $\beta$.}
\end{figure}

Case $iv$) As another case, bearing
Eq.~(\ref{memc}) in mind, defining $p_e=\beta p$,
$\rho_e=\rho+\alpha p$, where $\alpha$ and $\beta$ are unknown
constants that should meet the $\alpha+\beta=4$ condition, and
considering the $\dot{p}=\dot{p}_e=0$ case, one can follow the above
recipes to obtain

\begin{eqnarray}\label{fried6}
&&H^2=\frac{8\pi}{3}\rho_e,\\ &&3H^2+2\dot{H}=-8\pi p_e,\nonumber\\ &&q=\frac{1}{2}(1+\frac{3p_e}{\rho_e}),\nonumber\\
&&\dot{\rho}_e+3H(\rho_e+p_e)=0\rightarrow\rho_e=(\rho_e^0+\beta
p)(1+z)^3-\beta p,\nonumber
\end{eqnarray}

\noindent where $\rho_e^0$ is the integration constant.
As the previous cases, since the rhs of the two
first equations are nothing but the Einstein tensor meeting BI,
$\rho_e$ and $p_e$ should satisfy OCL (or equally the last line).
Indeed, since we have $\dot{p}_e=0$, the $\alpha+\beta=4$ constraint
is unavoidable only if we want KCL to be reduced to OCL (the last
line). We finally reach at

\begin{eqnarray}\label{fried62}
&&\rho=(\rho_e^0+\beta
p)(1+z)^3-4p,\\
&&w\equiv\frac{p}{\rho}=\frac{p}{(\rho_e^0+\beta
p)(1+z)^3-4p},\nonumber\\
&&w_e\equiv\frac{p_e}{\rho_e}=\frac{\beta p}{(\rho_e^0+\beta
p)(1+z)^3-\beta p}.\nonumber
\end{eqnarray}

\noindent As we know, the current accelerated universe ($z=0$)
implies $p_e<0$ and $w_e(z=0)\equiv w_0\leq-2/3$ (or equally
$\rho_e(z=0)>\frac{2}{3}(-p_e)$) \cite{roos}. Thus, the above
results admit a fluid with positive pressure whenever $\beta<0$.
In this manner, for $z\geq0$, if $\rho_e^0>\alpha p$, satisfied
when $\beta<\alpha w_0$ leading to $\beta<\frac{4w_0}{1+w_0}$ (or
equally $4<\alpha(1+w_0)$), then we have $\rho_e,\rho>0$. This
result indicates that we should have $w_0>-1$, in agreement with
some observations \cite{roos}, to meet the $\beta<0$ condition.
For this case, while, unlike $w$, $w_e$ is always negative, both
$w$ and $w_e$ approach zero for $z\gg1$. Using the above results,
we finally get
\begin{figure}[ht]
\centering
\includegraphics[width=2in, height=2in]{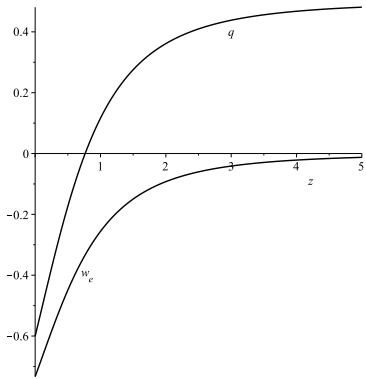}
\caption{\label{fig5} $w_e$ and $q$ versus $z$.}
\end{figure}
\begin{eqnarray}\label{fried63}
&&w=\frac{w_0}{\beta(w_0+1)(1+z)^3-4w_0},\nonumber\\
&&w_e=\frac{w_0}{(w_0+1)(1+z)^3-w_0},\\
&&q=\frac{(w_0+1)(1+z)^3+2w_0}{2[(w_0+1)(1+z)^3-w_0]},\nonumber
\end{eqnarray}

\noindent leading to $z_t=-1+(\frac{-2w_0}{1+w_0})^{\frac{1}{3}}$
for the transition redshift. The parameters $w$, $w_e$ and $q$
have been plotted in Figs.~(\ref{fig4}) and~(\ref{fig5}) for
$w_0=-0\cdot73$ \cite{roos}. It is also worthwhile reminding that
the current observational data on dark energy density and pressure
in fact give us the corresponding values of $\rho_e$ and $p_e$. As
it is apparent, there are some values of $\beta$ (and thus
$\alpha$) for which the maximum value of $w$ is at most equal to
$\frac{1}{3}$, signalling us to a baryonic source (since $\beta<0$
and $w>0$, we have $p>0$ and $\rho>0$). Therefore, $w_e$ and $q$
display proper behavior, meaning that it is mathematically enough
to only consider some parts of $T_{\mu\nu}$, represented by
$T_{\mu\nu}(e)$, to modify the standard cosmology
as~(\ref{fried6}) (or equally the Einstein field equations as
$G_{\mu\nu}=8\pi T_{\mu\nu}(e)$) which gives us a suitable
description for the current phase of the universe.

\section{Conclusion}

Mach principle states that geometry inherits its properties from
inside it, i.e. the geometrical information is related to the energy
of the source. Hence, it does not limit us to a certain information
for the energy source. This principle along with OCL
and BI are the backbone of Einstein theory, and
thus, standard cosmology, in full agreement with the thermodynamics
laws. Based on this theory, the Einstein ($G_{\mu\nu}$) and
energy-momentum ($T_{\mu\nu}$) tensors are in a direct relation as
$G_{\mu\nu}=8\pi T_{\mu\nu}$. Hence, by having the whole information
of the matter source ($T_{\mu\nu}$), one can find $G_{\mu\nu}$ and
then the spacetime metric which finally gives us the whole
information of geometry. It is also worthwhile
mentioning that the recent observation admits the break-down of OCL in
the current accelerated era \cite{prl}.

Besides, there are various approaches to gravity and cosmology in
which the Komar definition of energy is their foundation. Here, by
accepting this energy definition and using the thermodynamic
pressure definition, a relaxed energy-momentum conservation law
(KCL) has been obtained, helping us in getting
some modified energy-momentum tensors ($T_{\mu\nu}(e)$) satisfying
OCL. In continue, applying the thermodynamics laws to the apparent
horizon and attributing the Cai-Kim temperature to it, we could find
out some simple modified Friedmann equations.
Our results are in fact equal to modify the
Einstein field equations as $G_{\mu\nu}=8\pi T_{\mu\nu}(e)$.
Therefore, in agreement with the satisfaction of BI by the Einstein
tensor, $T_{\mu\nu}(e)$ is also meeting OCL. The
study shows that an appropriate choice of $T_{\mu\nu}(e)$
allows us to unify the dark sectors of cosmos into one
model. In fact, the analysis of the last case ($iv$) confirms that
if suitable $T_{\mu\nu}(e)$ is used instead of $T_{\mu\nu}$ in order to modify the
standard cosmology~(\ref{fried6}), then $T_{\mu\nu}(e)$ can
mathematically describe the current accelerated cosmos, while
$T_{\mu\nu}$ can display a baryonic source.

In summary, \textit{$i$) $T_{\mu\nu}(e)$, made of $T_{\mu\nu}$,
satisfies OCL (or equally $\nabla^{\mu}T_{\mu\nu}(e)=0$), and can
describe the current accelerated universe. $ii$) In this manner,
$T_{\mu\nu}$, satisfying KCL, can even show a baryonic behavior by
itself. These results may be translated as that the less
well-known or even unknown aspects of $T_{\mu\nu}$, represented by
$T_{\mu\nu}(e)$, are responsible for the current accelerated phase
of the universe. The origin and the emergence conditions of
$T_{\mu\nu}(e)$ may be found out by studying this part in the
framework of other physical theories such as quantum field
theory.}

\section*{Conflict of Interests}
The authors declare that there is no conflict of interest
regarding the publication of this paper.

\subsection*{Acknowledgments}

The authors would like to thank the referees for their valuable comments which helped to
improve the manuscript. The work of H. Moradpour has been supported financially by
Research Institute for Astronomy \& Astrophysics of Maragha
(RIAAM). J. P. Morais Gra\c ca and I. P. Lobo are supported by
Coordena\c c\~ao de Aperfei\c coamento de Pessoal de N\'ivel
Superior (CAPES).


\end{document}